\documentclass[12pt,preprint]{aastex}

\begin{document}


\title{ Accurate Realizations of the Ionized Gas in Galaxy Clusters:
Calibrating Feedback }


\author{Paul Bode and Jeremiah P. Ostriker} 
\affil{Department of Astrophysical Sciences, Peyton Hall, Princeton
University, Princeton, NJ 08544; 
bode@astro.princeton.edu, ostriker@princeton.edu}
\author{Jochen Weller}
\affil{University College London, Gower Street, London, WC1E 6BT, U.K.;
jochen.weller@ucl.ac.uk}
\author{Laurie Shaw}
\affil{Department of Physics, McGill University, Montreal, QC H3A 2T8;
lds@hep.physics.mcgill.ca}



\begin{abstract}
Using the full, three-dimensional potential of galaxy cluster halos
(drawn from an N-body simulation of the current, most favored
cosmology), the distribution of the X-ray
emitting gas is found by assuming a polytropic equation of state
and hydrostatic equilibrium, with constraints from conservation
of energy and pressure balance at the cluster boundary.
The resulting properties of the gas  for these
simulated redshift zero clusters (the temperature distribution,
mass--temperature
and luminosity--temperature relations, and the gas fraction)
are compared with observations in the X-ray of nearby clusters.
The observed properties are reproduced only under the
assumption that substantial energy
injection from non-gravitational sources has occurred.
Our model does not specify the source, but
star formation and
AGN may be capable of providing this energy,
which amounts to 3--5$\times 10^{-5}$ of the rest mass
in stars (assuming ten percent of the gas initially in the
cluster forms stars).
With the  method described here it is possible to
generate realistic X-ray and Sunyaev-Zel'dovich 
cluster maps and catalogs from N-body simulations, with 
the distributions of internal halo properties 
(and their trends with mass, location, and time)
taken into account.
\end{abstract}



\section{Introduction}   \label{sec:intro}

Current and upcoming surveys in a variety of wavelength bands 
will increase the number of well--observed 
clusters of galaxies by at least an order of magnitude, while 
probing to much higher redshifts than before.
Understanding the physical state of the intra-cluster medium (ICM)
will be essential to exploiting this new data.
In particular, it is necessary
to develop methods of accurately modeling 
the thermal state of the gas in clusters 
before one can extract cosmological
information from large surveys, which measure quantities
arising from that state.
For cluster-sized halos in a cosmological setting, 
the theoretical final distribution expected
from the gravitational collapse
of the dark matter (DM) is well understood
\citep{NavarroFW97,BullockKSSKKPD01,JingSuto02,PowerNJFWSSQ03,
ZhaoJMB03,NavarroHPJFWSSQ04,TasitsiomiKGK04,ReedGVGQSML05,
BartelmannDPBMMT05,DiemandZMSC05,ShawWOB06,LuMKW06}.
Measurements of the DM density profile in galaxy groups and
clusters agree well with this theoretical expectation
\citep{LewisBS03,DahleHS03,PrattArnaud05,PointecouteauAP05,
ComerfordMBS06,LokasWGMP06,RinesDiaferio06,ZekserWBBFIBPJC06,
MandelbaumSCBHB06,GastaldelloBHZBBM06,SchmidtAllen06z,SahaRW06}.
However, the hot intracluster gas in these systems does
not parallel the DM in either density or temperature
distribution.

Much progress has been made in understanding
the expected ICM distribution inside a standard DM halo
\citep[with the density profile showing a power law
cusp as in][or similar]{NavarroFW97,MooreQGSL99}.
\citet{MakinoSS98} gave an analytic expression for the density
of isothermal gas in hydrostatic equilibrium with a NFW
potential; this was soon extended to non-isothermal gas with
a polytropic equation of state 
\citep{SuotSM98,WuFN00,Loewenstein00,AscasibarYMS03},
and to triaxial halos \citep{LeeSuto03,WangFan06}.
The resulting gas profiles possess a finite density
core, and not a cusp as seen in the DM.  

The gross energetics of the gas do not
parallel that of the DM either.
Assuming that the gas energy comes solely from gravitational
collapse gives the self-similar scalings between mass $M$, 
luminosity $L$, and temperature $T$
of $M\propto T^{3/2}$ and $L\propto T^2$ \citep{Kaiser86,EkeNF98}.
However, these scalings do not agree with the observed relations,
leading \citet{Kaiser91} to propose that  non-gravitational energy
injection is important.  This idea has gained support from a
number of analytic investigations into polytropic ICM in DM
potentials \citep{BaloghBP99,SuotSM98,WuFN00,Loewenstein00,
TozziNorman01,KomatsuSeljak01,BabulBLP02,VoitBBB02,
DosSantosDore02,ShimizuKSS04,LapiCM05,AfshordiLS05,
SolanesMGS05}.
An additional departure from self-similarity can come from
star formation, which selectively removes gas with
short cooling times, low entropy, and low total energy,
leaving behind higher entropy material 
\citep{VoitBryan01,TozziNorman01,VoitBBB02,ScannapiecoOh04}.
Detailed computer simulations including star formation and
feedback are confirming the importance of non-gravitational
processes 
\citep[e.g.][and references therein]{BorganiFKPSTV05, 
EttoriDBM06,SijackiSpringel06,BorganDMCSDMTTT06,RomeoSPA06,MuanwongKT06,
Nagai06}.
It appears from the cited papers that both processes are at work:
low entropy gas is incorporated into stars, and energy and metals 
are added to the remaining gas via feedback processes.

Based on these advances in our understanding of clusters,
one can usefully combine prescriptions for gas physics
with analytic modeling of DM profiles, but this method
has limitations.
Since halos are formed by stochastic merging of subunits,
there is true scatter in halo concentration and inner slope
\citep{AvilareeseFKK99,Jing00,SubramanianCO00,BullockKSSKKPD01,
KlypinKBP01,FaltenbacherKNG05},
which can vary with time and halo mass
\citep{WechslerBPKD,Ricotti03,ZhaoJMB03,TasitsiomiKGK04,
SalvadorsoleMS05,ShawWOB06,WechslerZBKA06}.
Similar variations exist in halo triaxiality
\citep[e.g.][and references therein]{KasunEvrard05,
AllgoodFPKWFB06,HoBB06,RahmanKMFS06}
and substructure
\citep[e.g.][and references therein]{GillKGD04,ReedGQGSL05,ShawWOB06}.
The clustering of halos has been shown to depend upon
age or concentration, as well as mass 
\citep[][and references therein]{BerlindKBPSH06z}.
Thus mass, environment, and formation history all play significant 
roles in determining halo properties. 

In prior work,
\citet{OstrikerBB05} improved on the use of analytic DM potentials
by instead using the full three-dimensional potential of halos
drawn from N-body simulations, combining these detailed
3-D models with current modeling of the gas physics.
This procedure has the advantage of including 
the full distribution of halo concentration, as well as
halo triaxiality and substructure.  
By drawing on a large
N-body simulation volume (computationally much less costly than
a full hydrodynamic run), trends of internal halo properties with
mass, location, and time are all included, along 
with halo-halo correlations.
This procedure inevitably requires the utilization of certain
dimensionless parameters which--- since they derive from feedback
processes--- are difficult, at present, to determine from
{\it ab initio} computations.  The purpose of the present paper
is to tie down these parameters using observations
of X-ray clusters.  In particular, we 
apply the method of \citet{OstrikerBB05} to a set of halos
drawn from a large DM simulation of a cosmology in accord
with the WMAP 3-year data \citep{SpergelBDetc06z}.  The
resulting catalog is compared to X-ray observations of
nearby clusters.  The amount of energy input from
non-gravitational sources can significantly affect
gas properties, but with the proper amount--- consistent
with AGN activity---  this method can reproduce the
properties of the local cluster population.  

The procedures
used to create the simulated cluster sample are described
in \S\ref{sec:simclust}, these clusters are compared to
X-ray observations in \S\ref{sec:effects}, and we
discuss the implications in \S\ref{sec:discuss}.
One extension of this work is to use the gas properties
to compute the Sunyaev-Zel'dovich (SZ) effect; 
\citet{SehgalTHB06z} do this to make
available simulated 
large-area, sub-arcminute resolution microwave sky maps.

\section{The Simulated Cluster Catalog}   \label{sec:simclust}

\subsection{Dark Matter Halos}   \label{sec:dmhalos}

To produce a population of DM halos, we chose
cosmological parameters to match
the results from the 3-year WMAP data combined with
large-scale structure observations \citep{SpergelBDetc06z}.
The spatially flat LCDM model was used, with 
total matter density $\Omega_m=0.26$,
baryon density $\Omega_b=0.044$ 
(so the cosmic mean baryon fraction $f_c\approx 0.169$).
and cosmological constant $\Omega_\Lambda=0.74$.
Also, the Hubble constant
${\rm H}_{0}=72$km$s^{-1}$Mpc$^{-1}$
(i.e. $h=0.72={\rm H}_{0}/100$km$s^{-1}$Mpc$^{-1}$),
the primordial scalar spectral index $n_s=0.95$,
and the linear matter power spectrum amplitude $\sigma_{8}=0.77$. 
Concerning the N-body
simulation parameters, the number of particles $N=1024^3$ and
the box size $L=1000h^{-1}$Mpc, making the particle
mass $m_p=6.72\times 10^{10}h^{-1}M_\odot$;  the cubic spline
\citep[see][]{HernquistKatz89}
softening length $\epsilon=16.3h^{-1}$kpc.
The initial conditions for the $N$-body run were created
with the GRAFIC1 
code\footnote{This code is available at \url{http://web.mit.edu/edbert/}}
\citep{Bertschinger01}, with a few modifications.
Because the spherical Hanning filter employed in this code 
to isotropize small-scale structure also
significantly suppresses power on small scales, 
it was not used.
The linear DM transfer function at $z=0$ was calculated 
with the CMBFAST \citep{ZaldarriagaSeljak00} 
code\footnote{Available at \url{http://www.cmbfast.org/}}.
The DM growth factor was used to scale the resulting power spectrum
to the initial simulation redshift, chosen to be when the
density fluctuation amplitude on the grid scale is 10\%, $z=35.3$.
With these parameters, a cluster with mass 
$\sim 7\times 10^{14}h^{-1}M_\odot$ would contain $10^4$ particles;
a typical core radius for such a cluster would be $\sim 250h^{-1}$kpc,
or 15 times the particle softening length.

The simulation was run with the TPM 
code\footnote{Available at \url{http://www.astro.princeton.edu/$\sim$bode/TPM/}}
\citep{BodeOX00,BodeOstriker03},
with a couple of improvements over the publicly released 
version. 
Most variables in the improved code are double precision
(including particle positions and velocities), with the
main exception of accelerations and potentials, which are
still single precision.    Also, no lower limit was set to 
the parameter $B$ used in the TPM domain decomposition
\citep[see Eqn. 5 of][]{BodeOstriker03}. 
This means that at late times there will be more particles followed 
at full force resolution, leading to improved simulation of
the lowest mass objects;  at $z=0$ all cells with eight or more
particles were followed with trees.
The initial domain decomposition parameters were $A=1.9$ and $B=9.2$,  
the PM mesh contained $2048^3$ cells, and the maximum sub-box size was
256 cells.   At the end of the run, 54\% of the particles 
contained in 2\% of the total volume were being followed
with $5\times 10^6$ trees.

The standard friends-of-friends (FOF) halo finder
with linking length $b=0.2$ was run on the simulation volume at $z=0$,
identifying almost $2\times 10^6$ halos with 30 or more particles. 
The resulting mass function agrees well with the fitting
formula of \citet{WarrenAHT06};  the difference in the cumulative
mass function is less than 5\% over the range
$2\times 10^{12} \le M_{fof}/(h^{-1}M_\odot) \le 10^{14}$
and less than 15\% above this.
In what follows, only halos 
containing gas with temperatures above 1.5keV
will be included in our discussion.
We find only halos with $M_{500}>10^{13}h^{-1}M_\odot$
will ever meet this limit (though typically it takes over twice this mass).
$10^{13}h^{-1}M_\odot$ corresponds to 150 particles, so the
question arises as to whether or not there is sufficient numerical
resolution for such halos.  To test this, we compared the halos
described in this paper with a sample taken from a higher
resolution N-body simulation. 
This high resolution run has a box size $L=320h^{-1}$Mpc and a
spline softening length $\epsilon=3.2h^{-1}$kpc, so the
mass resolution is thus improved by a factor of 30 and the 
spatial resolution by a factor of 5;  otherwise
it was generated and evolved in exactly the same manner as the
simulation described above.   When applying the gas prescription
of the next section, we used a mesh cell size of
$l=9.60h^{-1}$kpc, a factor of 3.4 smaller than the standard
case (it was difficult to go to any finer mesh because of the
memory requirements of the resulting large computational arrays).
The maximum feedback model (described below) was assumed.
Table \ref{tab:restest} compares the mean and standard deviation
for several observables predicted in 
the standard and high resolution runs, using two mass bins
corresponding roughly to temperatures of 1.7 and 2 keV;
little difference is evident.  
A smaller cell size leads to a smaller volume around each
halo (for the same number of cells); this accounts for 
most of the differences seen in the two samples--- 
the mass resolution is relatively unimportant.  Increasing
the cell size to $l=12.80h^{-1}$kpc in the high resolution
run results in a distribution even closer to the standard run.

\subsection{The Gas Prescription}   \label{sec:gasp}

The gas distribution in each halo is calculated according
to the prescription of \citet{OstrikerBB05}.
Gas is placed in hydrostatic equilibrium
with the DM gravitational potential 
of the halo using a polytropic
equation of state.  Pressure balance with infalling
gas near the virial radius and energy conservation
determine the two constants required for the polytropic fit.  
Two important processes alter the gas energy.  Star formation
removes low entropy gas; we fix the conversion of gas into stars
at 10\%.  This leaves the most important free parameter, which
is the energy input into the cluster gas via feedback processes.
We will show that with a reasonable amount of feedback it is possible
to match X-ray observations of hot cluster gas.

In detail,
a cubic mesh enclosing the particles is placed around the
halo, with the cell size twice the particle spline softening
length, or $l=32.55h^{-1}$kpc.   The mass $m_p$ for each
particle is placed on the mesh using the cloud-in-cell
method, yielding the DM density, $\rho_{{\rm D}k}$, for each 
cell $k$.  The gravitational potential on the mesh, $\phi_k$,
is computed from the density as in a standard 
Particle-Mesh code, but with a nonperiodic FFT.  The 
center of the cluster is defined as the cell with the
lowest potential, $\phi_0=MIN(\phi_k)$.  The radii enclosing
various overdensities are calculated at this point, the
outermost being the virial radius, $r_{vir}$, enclosing the
overdensity expected from spherical tophat collapse, or
97 times the critical density at $z=0$ for the cosmology
used here
\citep[this is a change from][where overdensity 200 was used]{OstrikerBB05}.
The velocity of the cluster as a whole is taken to be the
mean velocity of the 125 particles closest to the cluster
center (or, for halos with fewer than 250 particles, the
innermost half).   Particle velocities are moved to the
rest frame of the cluster, and then the kinetic energy (KE)
of each particle is placed on the grid in the same manner
as the mass, yielding the KE per unit volume $\onehalf t_{{\rm D}k}$.

It is assumed that gas originally had the same
distribution as the DM, with density $f_c\rho_{{\rm D}k}$ and
KE $f_c \onehalf t_{{\rm D}k}$ ($f_c\equiv \Omega_b/\Omega_m$).
A certain amount of the gas mass, $M_\star$ 
(described below),
will have turned into stars;  this is presumably the most
bound material, so cells are ranked by binding energy
$\phi_k+\onehalf t_{{\rm D}k}$, and then cells are checked off until
the sum of the masses $f_c\rho_{{\rm D}k}l^3$ equals $M_\star$.
The initial mass $M_g$ and energy $E_g$ of the remaining gas are thus:
\begin{equation} \label{eqn:initmass}
M_g = \sum\limits_{k}  f_c  \rho_{{\rm D}k}  l^3
\hspace{1cm} ,
\end{equation}
\begin{equation} \label{eqn:initegas}
E_g = \sum\limits_{k}  f_c
\left\{ \phi_k\rho_{{\rm D}k}+\onehalf t_{{\rm D}k} \right\} l^3
\hspace{1cm} ,
\end{equation}
where the sum is over all cells inside $r_{vir}$ except
those marked off for star formation.
Also, the gas surface pressure $P_s$ on the cluster exerted by
surrounding material is estimated from the kinetic energy
in a buffer region nine cells (293$h^{-1}$kpc) thick 
outside of $r_{vir}$:
\begin{equation} \label{eqn:psurf}
P_s = N_{b}^{-1}\sum\limits_{k=1}^{N_{b}} 
\frac{f_c}{3} t_{{\rm D}k}
\hspace{1cm} ,
\end{equation}
where the sum is over the $N_{b}$ cells in the buffer region
$r_{vir}<r_k<r_{vir}+9l$.

Now suppose the gas is allowed to rearrange itself within
the DM potential such that it is in hydrostatic equilibrium
and has a polytropic equation of state with index $\Gamma=1.2$.
There is much support for such a model---
see the discussion and Fig. 1 in 
\citet[][comparing the polytropic model with a full, high resolution 
cosmological simulation by G. Bryan]{OstrikerBB05}, and
also \citet{AscasibarSYMG06}.
We will treat the gas as a tracer such that the potential,
set by the DM, does not change.
Defining
\begin{equation} \label{eqn:3dtheta}
\theta_k \equiv 1 + \frac{\Gamma-1}{(1+\delta_{rel})\Gamma}
  \frac{\rho_0}{P_0}\left(\phi_0-\phi_k\right)
\hspace{1cm} ,
\end{equation}
as in \citet{OstrikerBB05}, the resulting gas pressure $P$ and
density $\rho$ are given by
\begin{mathletters}
\begin{eqnarray} \label{eqn:p0rho0}
P(\vec{r_k}) = P_0 \theta_k^\frac{\Gamma}{\Gamma-1}
\hspace{1cm} , \\
\rho(\vec{r_k}) = \rho_0 \theta_k^\frac{1}{\Gamma-1}
\hspace{1cm} .
\end{eqnarray}
\end{mathletters}
Here $\delta_{rel}$ is a nonthermal component of pressure, assumed
to be proportional to thermal pressure such that the total
$P_{tot}=(1+\delta_{rel})P$.
To specify the final gas distribution given these assumptions, 
two quantities still need to be determined, namely
the pressure $P_0$ and density $\rho_0$ at the potential minimum.
This can be done with two equations of constraint, derived by
requiring conservation of energy and by matching the external
surface pressure, as follows.
For a given choice of  $P_0$ and $\rho_0$, the final radius
$r_f$ of the gas initially inside $r_{vir}$  can be found by summing
outwards from the cluster center until the initial mass $M_g$
is enclosed:
\begin{equation} \label{eqn:3dfm}
\sum\limits_{r_k<r_f} \rho_0 \theta_k^\frac{1}{\Gamma-1}l^3 = M_g
\hspace{1cm} .
\end{equation}
This implies that gas may expand or contract, changing the gas
fraction inside $r_{vir}$.  Assuming the external surface pressure
changes little with radius, there will be mechanical work done,
causing a change in energy proportional to the change in volume,
$\Delta E_p=(4\pi/3)(r_{vir}^3-r_f^3)P_s$.  The equation for
conservation of energy is thus
\begin{equation} \label{eqn:3dfe}
E_f=\sum\limits_{r_k<r_f} \left\{
\rho_0 \theta_k^\frac{1}{\Gamma-1} \phi_k
+ \frac{3}{2}(1+2\delta_{rel})P_0\theta_k^\frac{\Gamma}{\Gamma-1} \right\}l^3
= E_g + \Delta E_P + \epsilon_fM_\star c^2
\hspace{1cm} .
\end{equation}
The term $\epsilon_fM_\star c^2$ is feedback inferred from supernovae
and AGN, discussed in more detail below.
Matching the final surface pressure to the external pressure yields
the other equation of constraint:
\begin{equation} \label{eqn:3dfp}
(1+\delta_{rel})N_{b,f}^{-1} \sum\limits_{k=1}^{N_{b,f}}
P_0\theta_k^\frac{\Gamma}{\Gamma-1}
= P_s
\hspace{1cm} ,
\end{equation}
again summing in a buffer region $r_{f}<r<r_{f}+9l$.
\citep[Note the $\delta_{rel}$ terms were omitted in][]{OstrikerBB05}.
With Eqns. (\ref{eqn:3dfe}) and (\ref{eqn:3dfp}) it is possible
to iterate (e.g. with Newton-Raphson) to a solution for the final
gas density and pressure (or temperature) profile.

For this paper we will assume that at $z=0$ the initial 
(that is, inside $r_{vir}$ prior to any rearrangement)
star to gas
ratio is 10\%, in other words $f_\star=M_\star/M_g=0.10$, which 
implies $M_\star=(f_cM_{vir})f_\star/(1+f_\star)$.
This ratio agrees well with the value in nearby clusters measured
by \citet{LinMS03},
and is slightly lower than that measured by
\citet{VoevodkinVikhlinin}.
Star and black hole formation will return energy to the
remaining gas via supernovae and AGN activity--- writing this
energy as $\epsilon_fM_\star c^2$ of course assumes this
energy is proportional to stellar mass. 
This seems plausible:
the number of supernovae is expected to be proportional 
to the mass in stars, and the mass of central black holes
in spheriods is roughly proportional to the stellar mass
in these systems \citep{MerrittF01,KormendyG01}.
Improved observational constraints may alter this assumption.
The rough estimate
given in \citet{OstrikerBB05} is that $\epsilon_f\approx 3\times 10^{-6}$
for SN and $\epsilon_f\approx 4\times 10^{-5}$ for AGN, so we will
take $\epsilon_f=5\times 10^{-5}$ as the maximum case.  This is
roughly 3 keV per particle for the gas inside the virial radius,
which is at the high end of the plausible range.

\subsection{Cluster Temperature}   \label{sec:clusttemp}

To characterize the temperature of the gas we will use
$T_{ew}$, the X-ray emission-weighted $T$ inside a
projected radius of $R_{500}$.   The X-ray luminosity
is calculated using the cooling function $\Lambda(T)$ of
\citet{MallerBullock04}
for $T\le 10^8K$, and assuming $\Lambda(T)\propto T^{0.5}$ for
$T>10^8K$;  the metallicity is set to one third solar
\citep{BaumgartnerLHM05}.
\citet{MazzottaRMT04} showed that the
projected spectroscopic temperature of a thermally complex cluster
will in fact be lower than the emission-weighed temperature.
However, this difference will be more pronounced when computing
a single-temperature fit to a full hydrodynamic simulation
(containing shocks, cold fronts, and other short-lived structures) 
than it is
in our simple equilibrium models (which lack such local
inhomogeneities) because such features,
which increase the thermal complexity of the gas,
contribute to this
systematic bias \citep{MazzottaRMT04,KawaharaSKSSRD07}.  
In order to quantify this effect, we 
compared $T_{ew}$
with $T_{sp}$, the spectroscopic temperature measured
in the range $0.15R_{500}\le R\le R_{500}$. 
We used the 
code\footnote{Available at \url{http://hea-www.harvard.edu/$\sim$alexey/mixT}}
developed by \citet{Vikhlinin06} to 
compute $T_{sp}$, using the Chandra response function and
Galactic absorption $N_H=2\times 10^{20}\rm{cm}^{-2}$.
Note that simply excluding the center reduces the measured
$T_{sp}$ relative to $T_{ew}$, independent of spectral effects.
For the maximum feedback model,
we find ${\rm k}T_{sp}=0.94{\rm k}T_{ew}-0.06$keV, 
with little scatter;  in
other words, the two agree within 10\%.   Given this small
difference, we will use the conceptually simpler $T_{ew}$
unless stated otherwise.

\section{Effects of Feedback}   \label{sec:effects}

\subsection{Gas Temperature}   \label{sec:efftemp}

\begin{figure}
\plotone{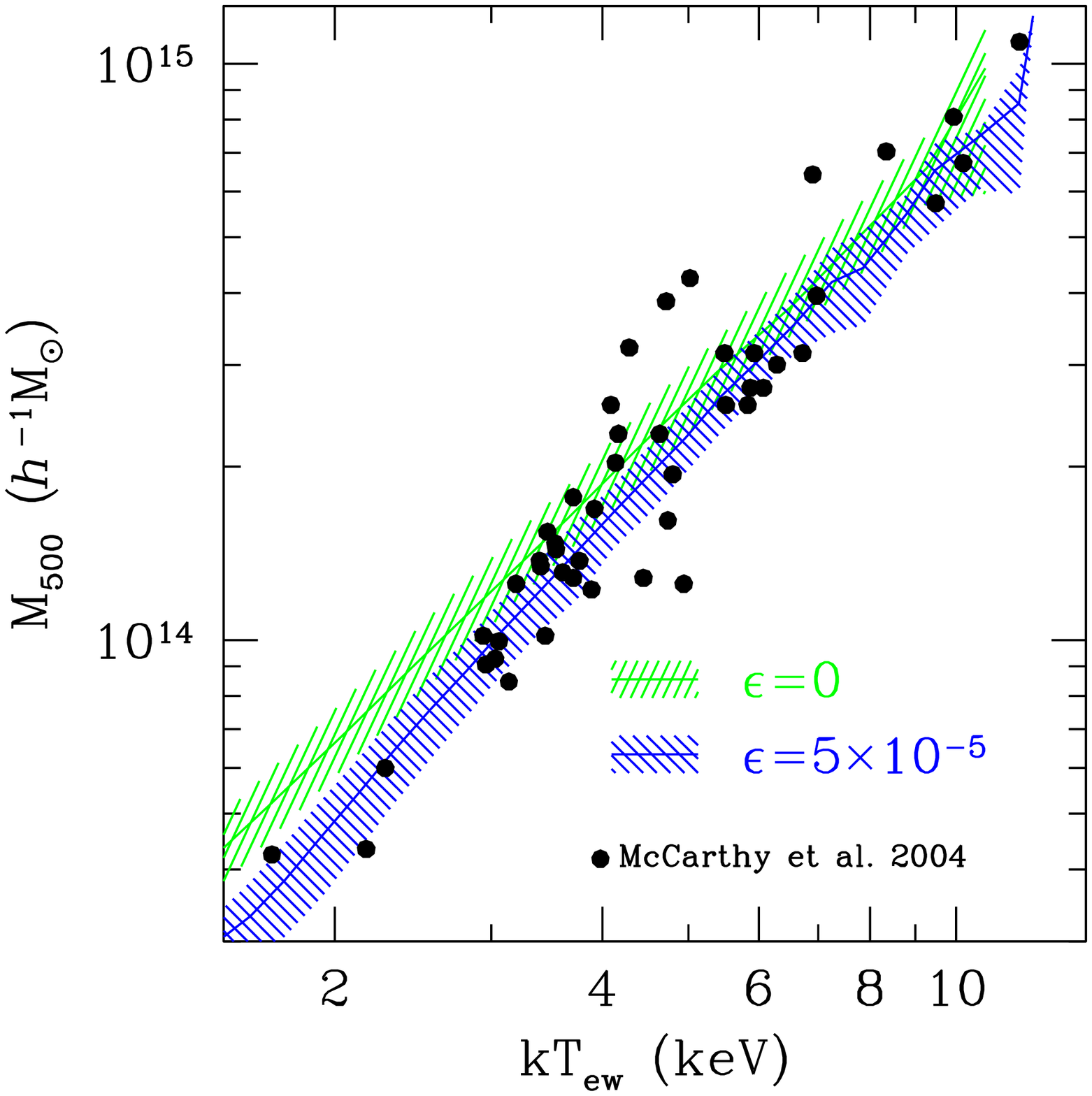}
\caption{ $M-T_{ew}$ relation for two values of the feedback
parameter. $T_{ew}$ is the emission-weighted temperature
from all material inside projected $R_{500}$.
Lines show the median value, and shaded regions enclose 68\%
of the clusters.
Filled circles are data described
in \citet{McCarthyBBPH04}, using only $z<0.06$ clusters.
\label{fig:tvsm} }
\end{figure}

X-ray surveys provide valuable information on the luminosity,
temperature, and mass of clusters.   
In this section we explore these properties as
derived using the method of the previous section.
The relationship between mass and temperature
is shown in Fig$.$\,\ref{fig:tvsm}.
The temperature is the emission-weighted $T_{ew}$ inside a 
projected radius $R_{500}$, and the mass is that contained in
a spherical overdensity of 500 times critical, $r_{500}$.   Lines show
the median mass at a given temperature, and the shaded
regions enclose 68\% of the clusters;
the cases of no feedback ($\epsilon_f=0$) and maximum feedback
($\epsilon_f=5\times 10^{-5}$) are shown.
Feedback has little effect for the hottest, most massive clusters:
the feedback energy is small compared to the binding 
energy of these clusters, and thus of little importance to the dynamical
state of the gas.  Feedback has a greater effect in less massive
clusters, making the gas somewhat hotter.  Still, at 
$M_{500}\approx 5\times 10^{13}h^{-1}M_\odot$, feedback increases
$T$ by only $\sim 33$\%.
One effect not apparent from the figure is that for masses below
$M_{500}\approx 3\times 10^{13}h^{-1}M_\odot$ the maximum feedback
can be enough to make the total gas energy positive, unbinding the gas
from the halo.  Thus our method produces no halos with temperatures
below about 1.5 keV in the maximum feedback case.
The data points shown in this figure 
are from \citet{McCarthyBBPH04}, who combined 
ASCA observations with the extended ROSAT
HIFLUGCS sample of \citet{ReiprichB02};  cluster cores were
not excluded when $T$ was determined.  Only those
clusters closer than $z<0.06$ are shown.
Below 4 keV, the feedback model provides a 
superior fit; both models are in agreement with the data
above this.  However, there is significantly more scatter
in the observed $M-T$ relation than is produced in our
model.   Cooling 
(which we neglect beyond that involved in star formation) 
will increase the scatter
in this relation \citep{McCarthyBBPH04,OHaraMBE06}. 
The existence of young systems which are out of dynamical equilibrium
can also broaden the observed $M-T$ relation, but the effects
may not be very pronounced \citep{OHaraMBE06}; we are to some extent
accounting for this, because
merging halos will have greater kinetic energy per particle
and thus a higher temperature than a relaxed halo of the same mass. 
We have not modeled observational error, which will also
of course add to any intrinsic scatter.

\begin{figure}
\plotone{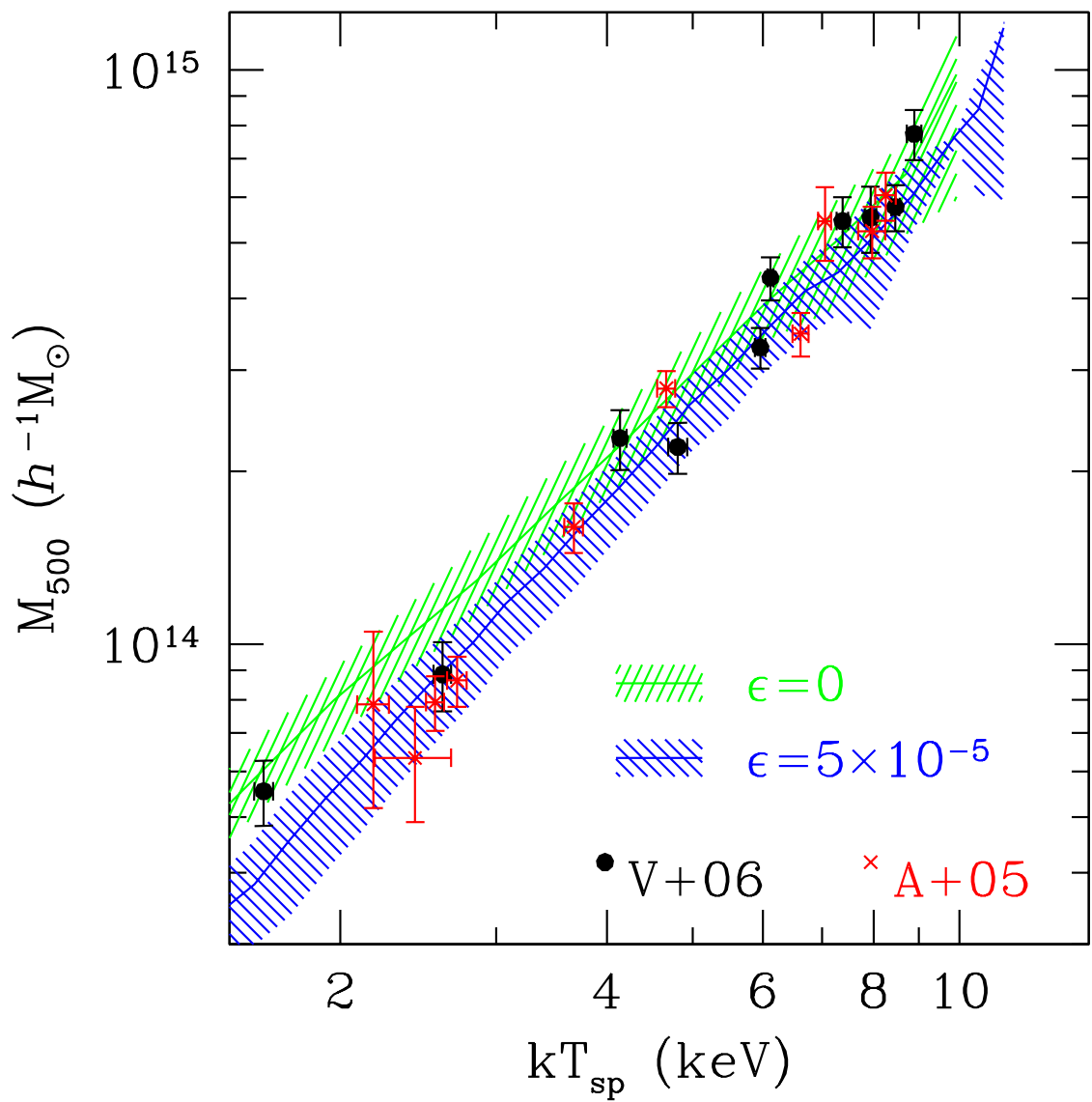}
\caption{ $M-T_{sp}$ relation for two values of the feedback
parameter.  $T_{sp}$ is the spectroscopic temperature
excluding the inner $0.15R_{500}$.
Lines show the median value, and shaded regions enclose 68\%
of the clusters.  Filled circles are from \citet{VikhlininKFJMMV06}
and crosses from \citet{ArnaudPP05}.
\label{fig:sptvsm} }
\end{figure}

Our predicted $M-T$ relations change little if we 
instead use the spectroscopic temperature, 
as shown in Fig$.$\,\ref{fig:sptvsm}.
Also shown are ten nearby, relaxed clusters observed with
XMM-Newton by \citet{ArnaudPP05}  and
ten relaxed clusters observed with
Chandra by \citet{VikhlininKFJMMV06}.
Both used spectroscopic temperatures, and excluded the
cores (although the radial range used to determine $T$
is slightly different).   These observations exhibit
considerably smaller scatter, lending credence to the idea
that differing dynamical states and cooling in cores will
increase the scatter in temperature at a given mass.
As was the case before, both models agree reasonably well
with the observed $M-T$ relation above 4 keV, but the
high feedback case is a better fit in the 2--3 keV range.
The $M-T$ relation can be well fit by the power law
\begin{equation} \label{eqn:mtpowlaw}
E(z)\frac{ M_{500} }{10^{14} h^{-1} M_\odot} = A
\left( \frac{ {\rm k}T}{\rm 5 keV} \right)^\alpha
\hspace{1cm} ,
\end{equation}
where $E(z)=H(z)/H_0$.
\citet{ArnaudPP05} found
$A=2.69\pm 0.10$ and $\alpha=1.71\pm 0.09$
for their sample, while  \citet{VikhlininKFJMMV06} obtained 
$A=2.89\pm 0.15$, $\alpha=1.58\pm 0.11$.
We first attempted to fit this relation to the simulated
halos with ordinary least-squares regression in the log-log
plane, but this was dominated by the more numerous low mass
halos and unduly affected by outliers.  Thus we instead
adopted the following procedure:  we divided the $x$-axis
into 20 logarithmically spaced bins, calculated the median in
each bin, and then found the best fit to these points.  This
in effect gives more massive clusters a higher weight in the
fitting; the resulting fits follow closely the median lines in
the figures.
As listed in Table \ref{tab:olsfits},
fitting to  all halos with ${\rm k}T_{sp}\ge 3$keV
in the zero feedback model gives
$A=3.12$ and $\alpha=1.49\pm 0.02$.
This follows the self-similar slope of 1.5, and gives
cooler clusters at a fixed mass than is observed.
In the maximum feedback case temperatures shift
to higher values, and the slope becomes
steeper: $A=2.56$ and $\alpha=1.62\pm 0.03$.
This slope agrees well with the observations, although
the normalization yields slightly hotter clusters at a given mass.
The formal error we obtain for $A$ is small (near one percent)
so in Table \ref{tab:olsfits} we give the $rms$ fractional
difference of the halos from the best-fit relation;  this scatter
reflects well the width of the shaded regions in the figures.
If the amount of feedback $\epsilon_f$
varied from cluster to cluster, then the scatter seen
would be larger.

\begin{figure}
\plotone{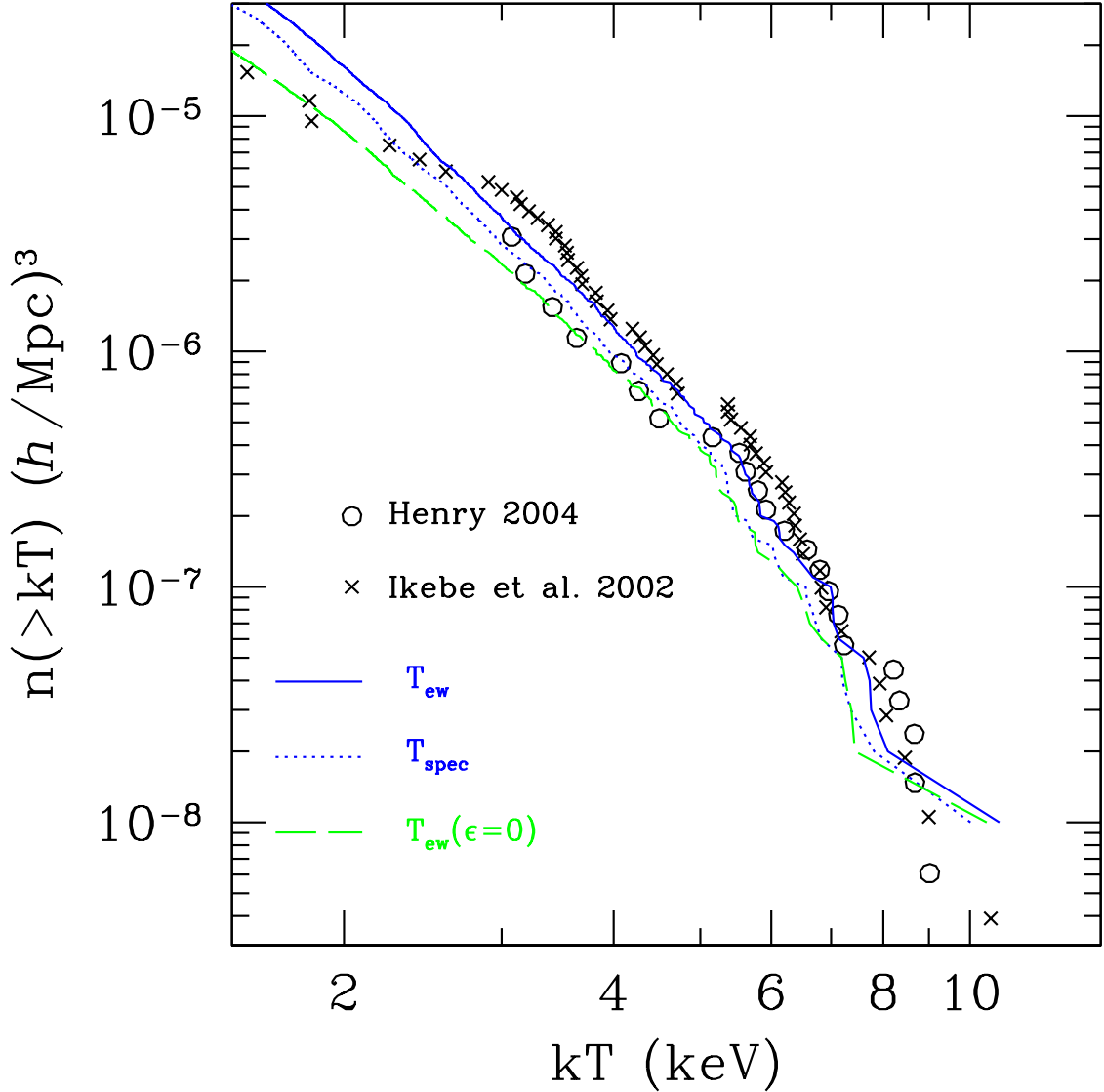}
\caption{ X-ray temperature function.  Crosses are from
\citet{IkebeRBTK02} (who excluded cluster cores), and
circles are from \citet{Henry00} (who kept them).
Lines are the volume limited
$z<0.2$ temperature function from simulated clusters:
{\it solid:} emission-weighted $T_{ew}$ from all material inside
projected $R_{500}$;  
{\it dotted:} spectroscopic $T_{spec}$ excluding the inner $0.15R_{500}$;
{\it dashed:} emission-weighted $T_{ew}$ with no feedback.
\label{fig:noft} }
\end{figure}

Another method of characterizing nearby clusters is
the temperature function, which does not require a
mass determination.  The distribution of cluster
temperatures is sensitively dependent on the cosmological
model; in \citet{OstrikerBB05}, which used the first-year
WMAP power spectrum amplitude, the fit to observations was
inadequate.
Shown in 
Fig$.$\,\ref{fig:noft} are two measurements
of the cumulative temperature function, with different
methods of determining the cluster temperature:
\citet{IkebeRBTK02} excluded cluster cores when fitting
for $T$, while \citet{Henry04} did not.  
For purposes of comparison, we took from the simulation
a ``light cone'' covering one octant of the sky out to
$z=0.2$.   
This covers the redshift range used in the observations,
although they are not volume limited.
During the simulation, the matter distribution in a
series of thin shells was saved;  the radius of each shell
corresponds to the light travel time from a
$z=0$ observer sitting at the origin of the box, and
its width corresponds to the time interval between shells.
Thus a volume-limited mass distribution, including
time evolution, is obtained.  Locating halos and adding
gas was done in the same manner as before.  To compute
the star/gas ratio at $z>0$, 
the star formation rate was assumed to follow a delayed
exponential model \citep[Eqn. 1 of][]{NagamineOFC06},
with decay time $\tau=1$Gyr.   Both $T_{ew}$ and $T_{sp}$ 
are shown for the $\epsilon_f=5\times 10^{-5}$ feedback model in 
Fig$.$\,\ref{fig:noft}.  Because our cosmological
parameters were chosen in part to match large-scale structure
observations, and the simulated $M_{500}-T$ relation matches
that of nearby clusters, it is not surprising that our simulated
temperature function is a reasonable fit to the observed one.
Our $T_{ew}$, which includes the core, gives a higher abundance 
in the $3-6$ keV range than the \citet{Henry04} data, while
our $T_{sp}$, excluding the core, instead gives a lower
abundance than \citet{IkebeRBTK02}.
The zero feedback model appears to give temperatures that
are too low;  thus based simply on $T$, it appears that
some non-gravitational energy input is required to 
explain the properties of existing clusters.
The model with feedback and 
WMAP 3-year cosmological parameters now provides
a good fit to the observed temperature function.

\subsection{Gas Density}   \label{sec:effden}

Other cluster observables are more dependent on
the gas density, most notably X-ray luminosity.
The top panel of Fig$.$\,\ref{fig:lxandf} 
shows the $L_x-T$ relation of our simulated catalog
for three values of feedback,
again with medians shown as lines and shaded regions
enclosing 68\% of the clusters;
here $L_x$ is the bolometric luminosity inside a 
projected radius of $R_{500}$.  
Unlike the $M-T$
relation, feedback produces a significant
change in $L_x$ at a given temperature because
the Bremsstrahlung emission is proportional to the square of the
gas density, but otherwise there are similar trends seen.
Again the effect of feedback is
less important in the most massive clusters, 
where gravitational binding energy dominates.
Also, the scatter for a given amount of feedback is
much smaller than that observed in nearby clusters
\citep[the data points are again from][with $z<0.06$]{McCarthyBBPH04}.
The zero feedback and $\epsilon_f=5\times 10^{-5}$ models bracket
the range of luminosities seen in nearby clusters.  An 
intermediate model with $\epsilon_f=3\times 10^{-5}$
is also shown in Fig$.$\,\ref{fig:lxandf};  it appears
this is an insufficient amount to explain the lowest
luminosity clusters.  Fitting a power law 
$L_x/(10^{44}h^{-2}{\rm erg}\,s^{-1})=A({\rm k}T_{sp}/5{\rm keV})^\alpha$
to all our halos with ${\rm k}T_{sp}\ge 3$keV yields 
$\alpha=1.66\pm 0.03$ in the zero feedback case, not as steep as
the self-similar expectations of of $L_x\propto T^2$.
Including feedback steepens this relation considerably
to $\alpha=2.51\pm 0.08$, more in line with the observed
value \citep[e.g.][]{ArnaudEvrard99,ReiprichB02,IkebeRBTK02}.
The exact value of the slope we find
depends on the lower temperature limit used.

A more direct probe of gas density is the gas fraction
within a given radius.   It appears that
the gas fraction increases with increasing radius, and 
higher temperature clusters have a higher gas fraction
as well 
\citep{DavidJF95,ArnaudEvrard99,MohrME99,VikhlininKFJMMV06}.
The gas fraction from our model is shown in the bottom
panel of Fig$.$\,\ref{fig:lxandf}, with
$f_g$ defined as the fraction of the total
mass inside a spherical radius $r_{500}$ enclosing
an overdensity 500 times critical;  
for the purpose of computing the total mass, we assumed that
stellar mass followed the same radial profile as the dark matter.
The model curves display the type of
behavior one might expect based on the top panel:
models with feedback show significantly lower gas 
fractions (i.e. lower densities at a given $T$), with the effect
being most pronounced for the lowest mass clusters.
Recent observations of the gas fraction by
\citet{VikhlininKFJMMV06} are shown in the figure,
as well as two clusters from  \citet{GastaldelloBHZBBM06},
where for these points the temperature was derived from $M_{500}$ 
using the $M-T$ relation of 
\citet[][note they use a spectroscopic temperature, and exclude
the central region when finding $T$]{VikhlininKFJMMV06}.
Also shown as a dashed line is the best fit $f_g-T$ relation found by 
\citet{MohrME99} using 45 ROSAT clusters and temperatures 
taken from the literature.  
Here again,
some feedback is required to bring the models in
agreement with observed gas fractions, but again the
spread in any given model is too small to fit all
observed values of $f_g$.

Also shown as a dotted line in the figure is the mean
gas fraction from our cosmological model, after turning
enough gas into stars to make the global star/gas ratio
ten percent, or $\bar{f_g}=f_c\cdot(1-f_\star/(1+f_\star))$.
Without any feedback, the baryon fraction inside $r_{500}$
will reflect the cosmic mean value, but energy input
drives this fraction lower, particularly for smaller
clusters.  
We find the gas fraction increases with
radius, so at overdensities higher than 500 this
discrepancy will be even greater.  This raises the question
of how far out one must go before the cluster contains
a fair sample of the cosmic mass budget.
Fig$.$\,\ref{fig:gfrac} shows the median gas fraction
inside the virial radius as a function of temperature
(still using $T_{ew}$ inside $R_{500}$);
the virial overdensity is 97 times critical for our chosen cosmology.
Even at this radius, it is only for the most massive
clusters (${\rm k}T_{ew}\gtrsim 6$keV) that the baryon
fraction reaches the cosmic mean;  in smaller clusters
feedback causes the gas to expand, reducing the gas fraction
by many tens of percent.  Attempts to determine the cosmic
baryon density from clusters will need to take this effect
into account.   

Fig$.$\,\ref{fig:gfrac} also demonstrates how the
median gas fraction changes in the maximum feedback case
if we also add a relativistic component with $\delta_{rel}=0.20$
(there is little change in the spread around the median).
With this component, a lower temperature is required to
achieve pressure balance at a given density.
This model behaves like the no feedback case at higher
masses, and like an intermediate feedback case at
lower masses.  This behavior also holds for 
all the other relationships
($M_{500}-T_{ew}, L_x-T_{ew}$, etc.) explored in this paper;
thus it seems a relativistic component will have little
effect on thermal cluster observables.  
While our 
implementation is quite simplified, similar results
were found using full hydrodynamic simulations 
by \citet{PfrommerSJE06z}:  they found that including
the effects of cosmic rays caused only small changes
in the gas fraction and integrated SZ signal
(they found a larger change in $L_x$, but this was
related to cooling cores, which we do not implement). 

To summarize this section, we find that a
WMAP 3-year cosmological model coupled with a
feedback parameter of $\epsilon_f=3-5\times 10^{-5}$
(which corresponds to an input energy of roughly
2-3 keV per baryon) provides a good fit to
the extant X-ray observations of hot gas in clusters.

\section{Discussion}   \label{sec:discuss}

In this paper we
have presented a method for determining the gas
distribution inside a fully three-dimensional potential;
this method assumes hydrostatic equilibrium and a polytropic
equation of state, and also that the original gas energy
per unit mass equals that of the DM.
We then applied this method
to a $z=0$ catalog of DM cluster halos drawn from N-body simulations,
and compared the resulting  ICM distributions to
observations of nearby clusters.   The main result is that
this simple gas prescription can reproduce many of the observed
bulk properties of the ICM, including the temperature 
distribution and the relationships between temperature and
mass, X-ray luminosity, or gas fraction.
The main drawback is that
in nearby X-ray clusters there is significantly more 
scatter seen in these relationships than is produced in our
method.   This could be due to a number of factors, 
including cooling cores, our assumption of
hydrostatic equilibrium, and observational errors.

The advantages of using this type of model for the ICM  are clear.
It is possible to simulate a large volume 
with a N-body code at much smaller computational
cost than is required for a full 
hydrodynamical treatment.
When adding gas with the method described here,
the distributions of
internal halo properties--- concentration, triaxiality,
substructure, etc.--- are taken into account, as are their
trends with mass, location, and time, plus any 
alignments and correlations between halos.
However, it should be noted that this approach does not 
account for the dynamical effects of a baryonic
component.  Results from halo formation simulations 
demonstrate that including a dissipational gas component 
can alter the radial profile of the DM halo
\citep{GnedinKKN04,LinJMGM06} and its ellipticity
\citep{KazantzidisKZANM04}, which would in turn alter
the ICM distribution. 

\begin{figure}
\plotone{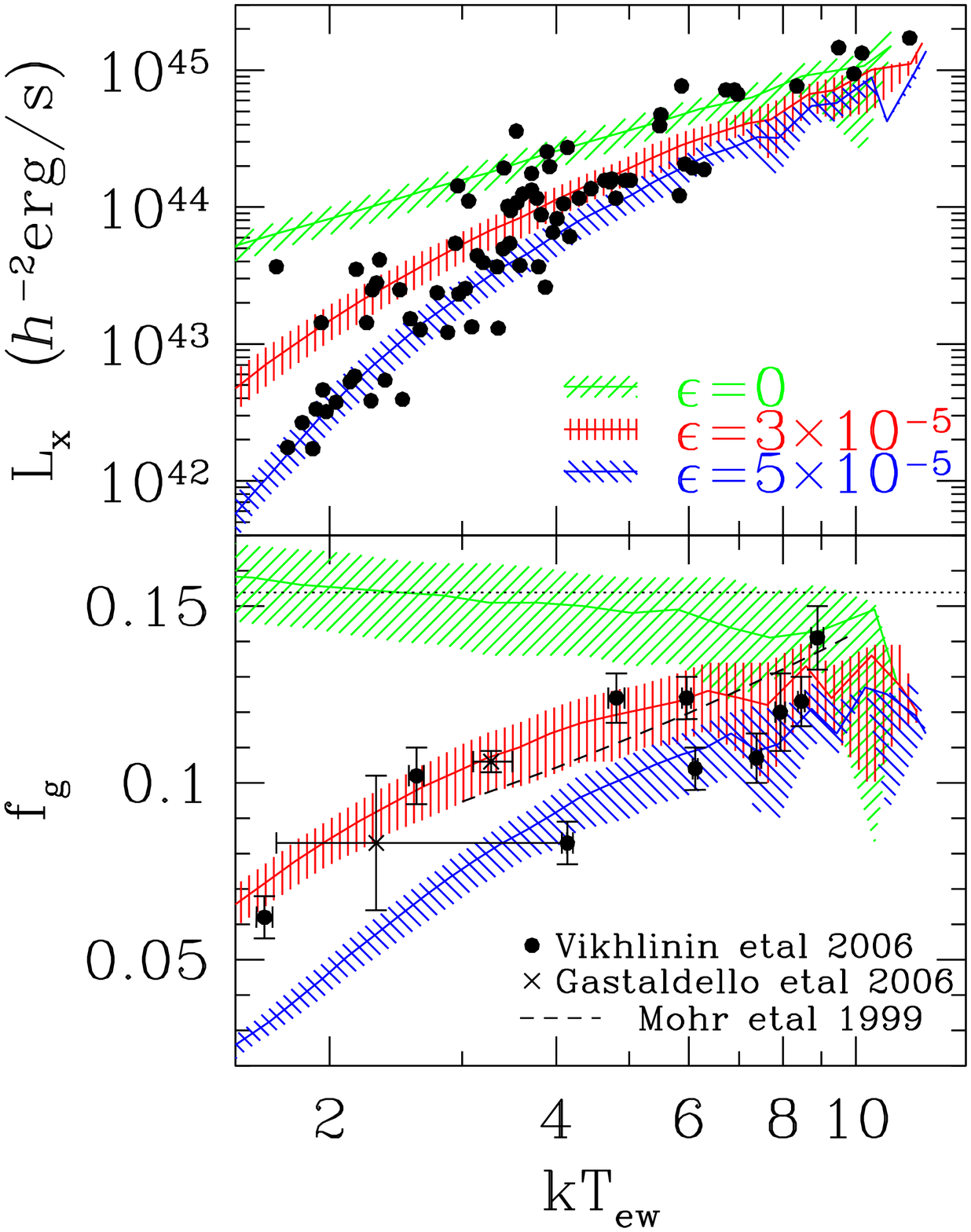}
\caption{ Top panel: $L_x-T$ relation for three values of
the feedback parameter.
For the simulated $z=0$ clusters, the lines are the median
and the shaded regions enclose 68\% of the clusters.
Filled circles are data described
in \citet{McCarthyBBPH04}, using only $z<0.06$ clusters.
Bottom panel: gas fraction inside $r_{500}$.
Points with error bars are data from 
\citet{VikhlininKFJMMV06} and \citet{GastaldelloBHZBBM06},
and the dashed line is the best fit to 
45 ROSAT clusters by \citet{MohrME99}.
The dotted line is the cosmic mean adjusted
to make the global star/gas ratio 10\%.
\label{fig:lxandf} }
\end{figure}

\begin{figure}
\plotone{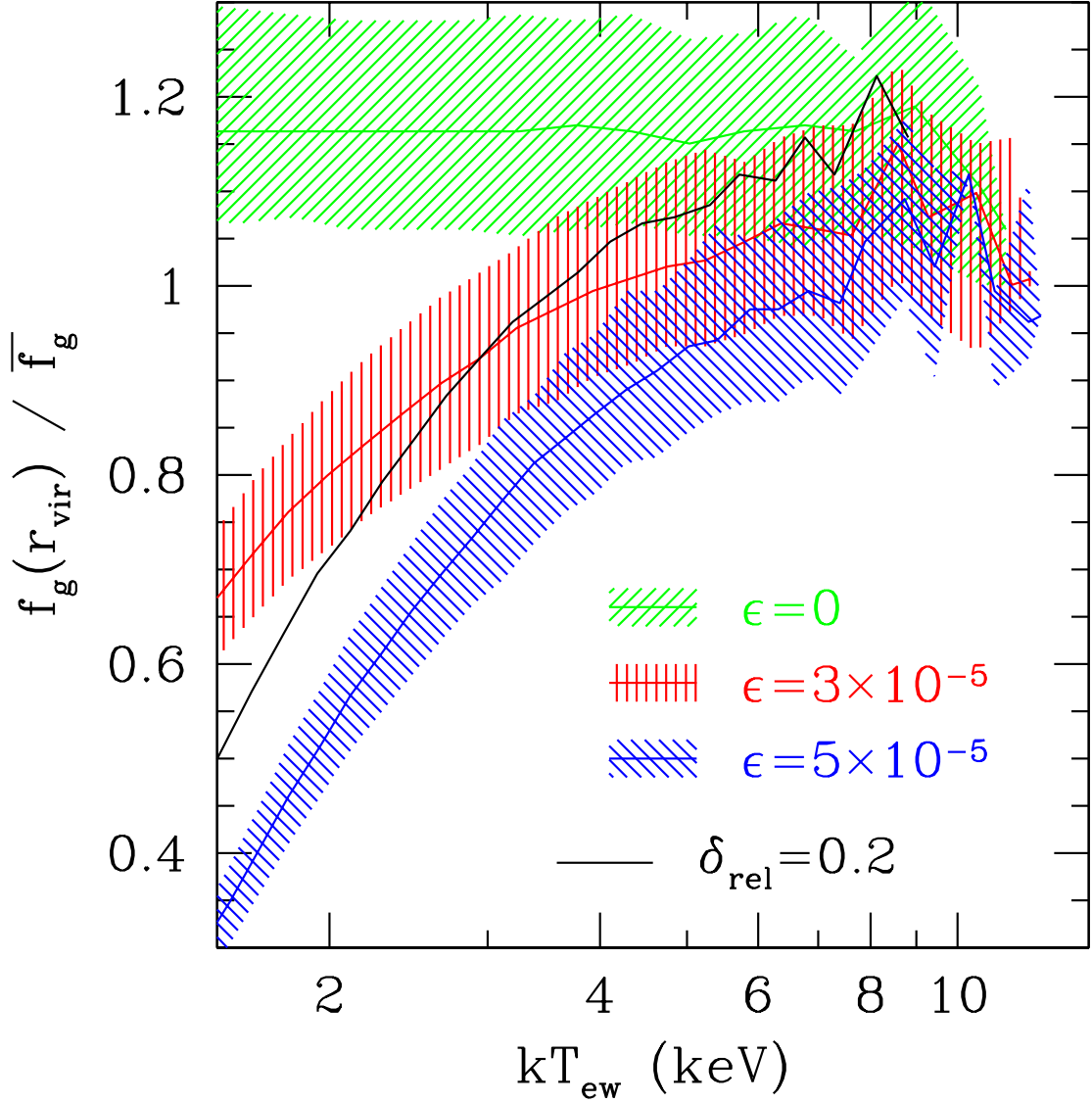}
\caption{ Gas fraction inside the virial radius $r_{vir}$,
normalized to the cosmic mean value (adjusted for star formation),
as a function of temperature.
Lines are the median
and shaded regions enclose 68\% of the clusters.
The line without shading is the median for 
$\epsilon_f=5\times 10^{-5}$ and
$\delta_{rel}=0.20$. 
\label{fig:gfrac} }
\end{figure}

A second result is that a significant amount of  non-gravitational 
energy input is required to reproduce the properties
of nearby clusters.   
Many other investigators have reached the same conclusion
\citep[e.g.][]{Kaiser91,BaloghBP99,SuotSM98,WuFN00,Loewenstein00,
TozziNorman01,KomatsuSeljak01,BabulBLP02,VoitBBB02,
DosSantosDore02,ShimizuKSS04,LapiCM05,AfshordiLS05, 
SolanesMGS05}. 
In this paper we do not try to find the sources of this input.
Instead, we simply ascertain what level of feedback would
produce the observed relations among cluster observables.
Processes involved in star formation provide some
of this energy, but not enough.
The most likely source for extra energy
is AGN, which can conceivably deliver enough feedback 
to explain the temperature, X-ray luminosity, and gas fraction 
distributions of local clusters.   However, there is little
margin for error:  if  we have overestimated the amount
of star formation, 
the black hole to stellar mass
ratio is less than 0.0013, and/or the amount of energy returned
to the gas by black hole formation
is less than 3\% of the black hole rest mass,
then the required energy of 2 to 3 kev per particle
will not be produced.
This conclusion differs somewhat from that of \citet{McCarthyBB06},
who find that AGN heating is an implausible (though not impossible)
explanation of cluster gas fractions;  they calculate that 
in order to reduce the gas fraction within $r_{500}$ to the observed level
it would take 10 keV per particle in the gas observed within
$r_{500}$ (roughly 0.12 $M_{500}$).  Rescaling this value to 
our normalization (energy per particle of the gas mass initially
inside $r_{vir}$, or $f_cM_{vir}$), makes this a required energy of
roughly 4 keV per particle.
The reason we require less energy is in part due to
a different initial state: examining
the lower panel of Fig$.$\,\ref{fig:lxandf}, by just
accounting for star formation but not including any
feedback, one can see our method leads to a gas fraction inside  $r_{500}$
already lower than the cosmic mean, 
while \citet{McCarthyBB06} started with
a state in which the gas fraction equals the cosmic mean. 
(Although in any case they argue that efficiencies of AGN outbursts
are $\sim10^{-3}$, not $\sim10^{-1}$).
Running hydrodynamic simulations without radiation or feedback,
\citet{CrainEFJMNP06z} found the baryon fraction 
inside $r_{500}$ of 90\%, which agrees well with our method
at higher masses.  However, \citet{CrainEFJMNP06z} also find
this fraction still holds at lower masses, and that the baryon
fraction is still 90\% at the virial radius (again with no
feedback), while  we find a higher fraction in both of these instances.
Note that if we had instead started with an initial state consistent
with a 90\% baryon fraction, then we would require a lower
amount of feedback to reproduce observed gas fractions.

\begin{figure}
\plotone{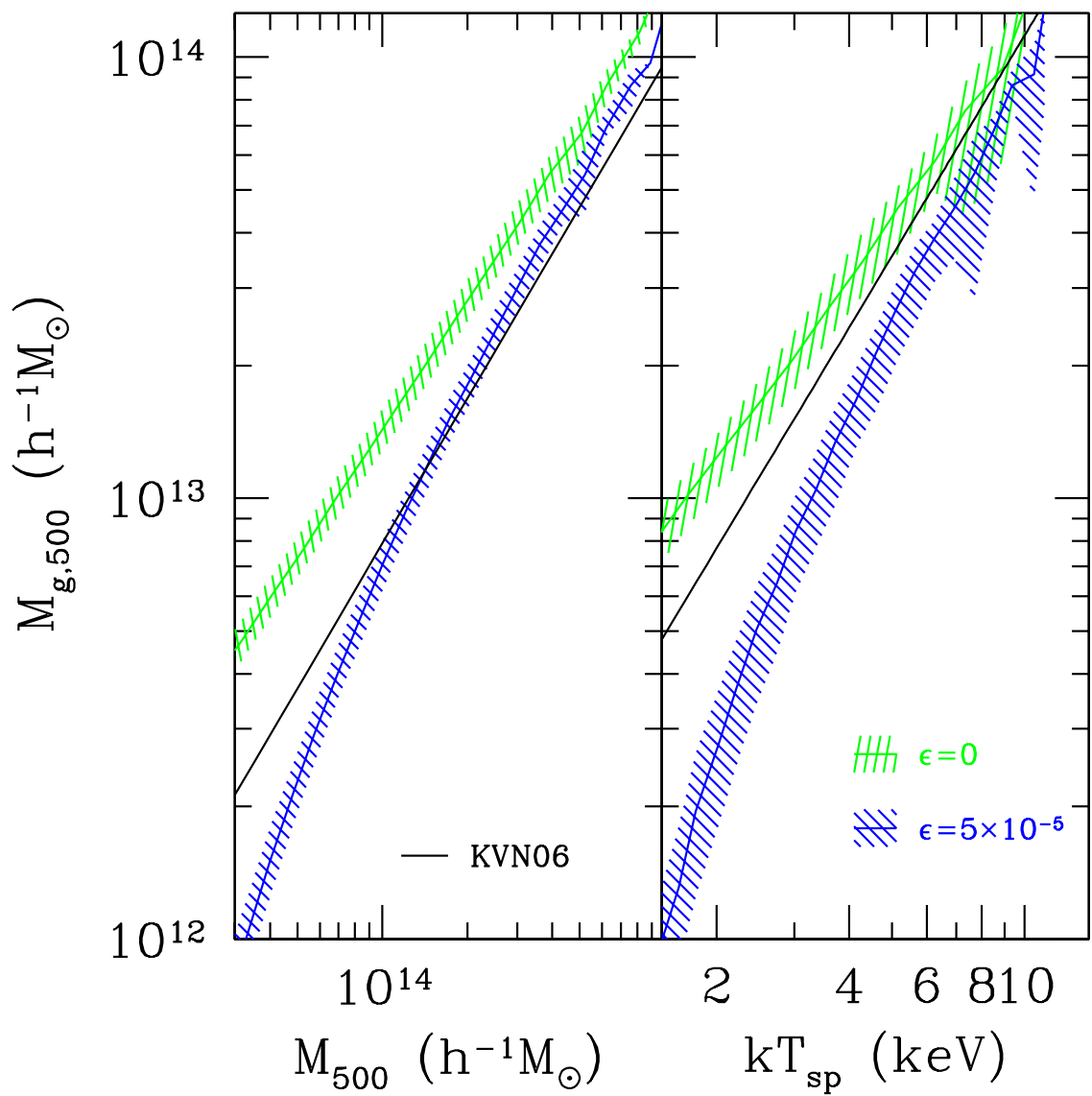}
\caption{Gas mass inside $r_{500}$ as a function of $M_{500}$ (left)
and ${\rm k}T_{sp}$ (right).  The straight lines are the
full hydrodynamic simulation result from \citet{KravtsovVN06}.
Lines are the median
and shaded regions enclose 68\% of the clusters.
\label{fig:sims} }
\end{figure}

It is useful in this context to compare with
recent hydrodynamical simulations which include feedback.
The left panel of Fig$.$\,\ref{fig:sims}
relates gas to total mass inside $r_{500}$;
the solid line shows the best-fit power law
relation at $z=0$ found from adaptive 
refinement simulations by \citet{KravtsovVN06}.
With no feedback we find a similar slope (slightly larger
than unity), but
for a given halo mass there is a higher gas
fraction than in the simulations.  
Adding feedback reduces this discrepancy at
higher masses, but at lower masses our maximum 
feedback case instead has lower gas fractions.
As \citet{KravtsovVN06} did not attempt to 
include AGN feedback it is not surprising that
an intermediate amount of feedback would provide
the best match.   We do not match the simulations
when it comes to the total mass-temperature relation, however.
For Eqn. (\ref{eqn:mtpowlaw}), \citet{KravtsovVN06} find
$A=3.85\pm 0.19$ and $\alpha=1.524\pm 0.07$,
while \citet{KaydSABLPST06z} with an SPH simulation find
$A=4.47\pm 0.19$ and $\alpha=1.76\pm 0.07$.
We obtain a similar slope ($\alpha=1.62$), 
but more importantly we
find a different normalization---
that is, we find hotter temperatures at a given mass,
as would be expected because we
included a higher level of feedback, 

The near future will see a number of surveys that select
clusters of galaxies via their Sunyaev-Zel'dovich (SZ) decrement, which is
proportional to the gas pressure in the cluster integrated along the
line of sight. Currently the 
Sunyaev-Zel'dovich Array \citep[SZA;][]{Carlstrom:00} and the 
Arc Minute Microkelvin Imager \citep[AMI;][]{Kneissl:01} are equipped to
perform such a search on tens of square degrees on the sky.   However,
the Atacama Cosmology Telescope (ACT), the South Pole Telescope (SPT), 
APEX, and ultimately the Planck surveyor will scan thousands of square
degrees on the sky in the radio \citep{Ruhl:04,Fowler:04,Gusten:06,Planck}.  
These surveys will detect thousands of clusters;
for example, the SPT
will scan 4,000 $deg^2$ on the sky and observe of the order of $6,000$
clusters of galaxies (this is for a flux
sensitivity of about $1.5$mJy at the 4-sigma detection threshold
with a 1 arcmin beam operating at 150 GHz, and assumes $\sigma_8 = 0.75$). 
This translates to a
limiting mass of $M_{\rm lim}\approx 10^{14.2}h^{-1}M_\odot$ if one
assumes the clusters are in hydrostatic equilibrium.
\citet{SehgalTHB06z} found a similar limit for a 90\% complete
cluster sample from ACT.

The redshift distribution of clusters is very sensitive to the amplitude 
and growth of linear perturbations, and hence to cosmological parameters 
\citep{Holder:01,Haiman:01,Weller:02,Battye:03,Majumdar:04, Younger:06}. 
However, in order to exploit SZ cluster number
counts one is required to understand the selection function of these
surveys.  This is most easily accomplished
in terms of the flux decrement, which depends on
the system temperature, exposure time, band-width, and efficiency
\citep{Battye:05}. In order to obtain cosmological constraints,
it is necessary to convert the observables
into a limiting mass of the
survey.  There are two approaches to obtain this mass limit. 
One is to start with the assumption that all clusters are 
spherical and in hydrostatic equilibrium,
and then include some nuisance parameters
to allow for deviations from this assumption
\citep{Verde:02,Battye:03,Younger:06}.  
Another approach is to use a very general
parameterization of the mass--observable relation, which in its most
general form could easily introduce forty unknown parameters
\citep{Hu:03a,Lima:05}.  
Currently there is little data to constrain the free parameters
in either approach.  In future one can
exploit the SZ cluster observations themselves, to 
self-calibrate these free parameters
\citep{Hu:03a,Lima:05,Majumdar:04,Battye:03}. 
However if one employs the most general
parameterization, little power is left in the surveys to constrain
cosmological parameters \citep{Hu:03a,Lima:05}. 
Another possibility would be
to use complementary observations, such as weak lensing to
cross-calibrate the mass-observable relation
\citep{Majumdar:04,DES,Sealfon:06}. 
A useful approach would be to
have a physical parameterization of the mass-observable relation with
some prior probability on the free parameters and then self-calibrate
the SZ surveys for these parameters \citep{Younger:06}. 
However in order to
obtain this prior knowledge we can not yet rely on observations
because currently they are sparse, 
and in the near future observations will not resolve
clusters because the beams of the instruments are typically 
larger than 1 arcmin. We
hence require simulations to explore the scatter in the
mass-observable relation; in order to obtain realistic results, a large
representative sample of galaxy clusters is required.

\begin{figure}
\plotone{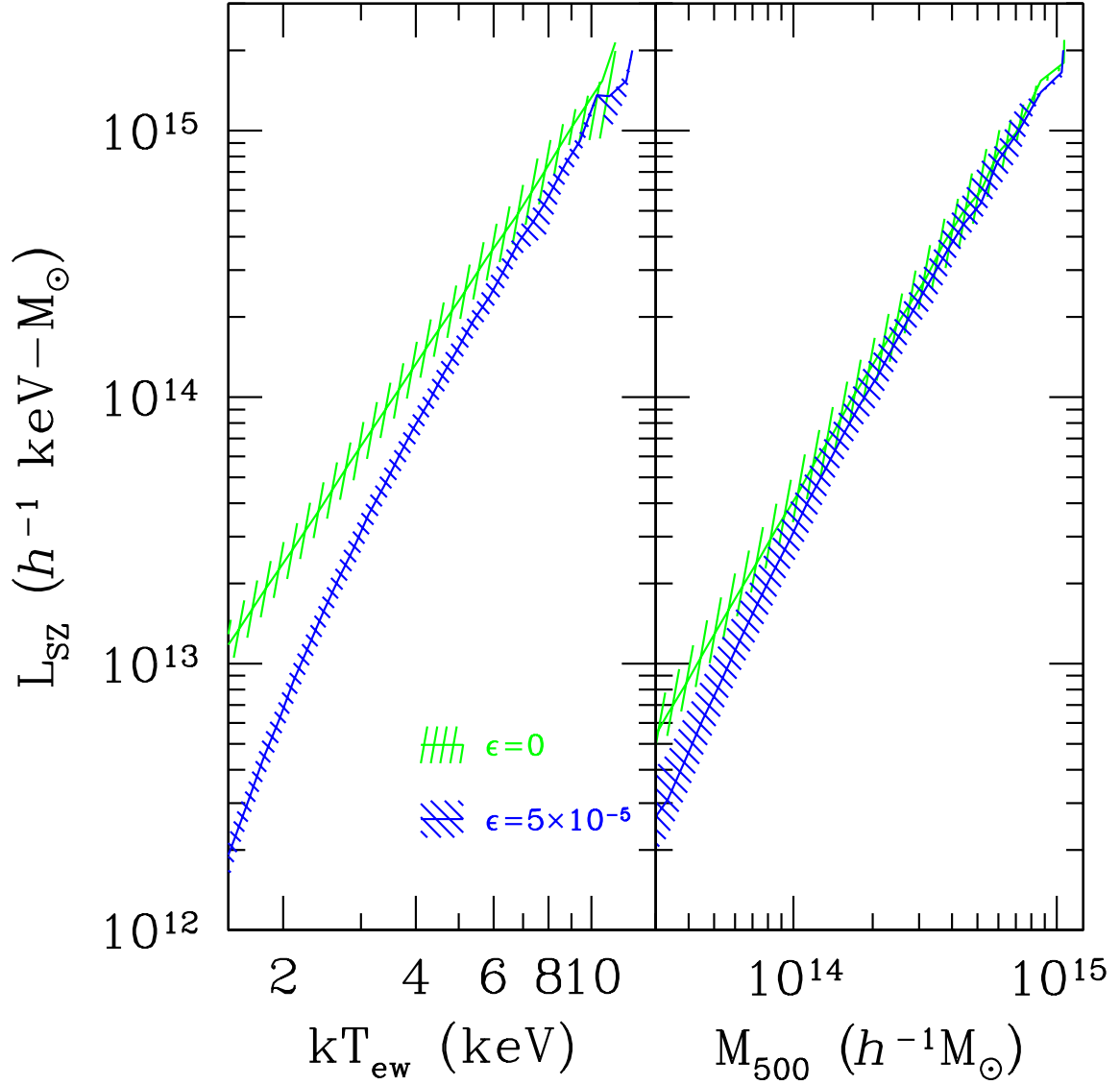}
\caption{Integrated SZ decrement,
$L_{SZ} = \int dA \int \rho {\rm k}T dl$, 
as a function of temperature (left) and mass (right),
for two values of the feedback parameter.
Lines show the median value, and shaded regions enclose 68\%
of the clusters.
\label{fig:sz} }
\end{figure}

\citet{SehgalTHB06z} have already applied the method of this
paper to a full light-cone N-body output (out to $z=3$)
in order to generate and make publicly available
large-area, sub-arcminute resolution microwave sky maps.
We intend to provide
a detailed analysis of the mass-observable relation in a forthcoming
paper, and will give here only a rough qualitative discussion.
In particular,
we can use our $z=0$ simulated catalog to explore how
the amount of thermal energy in the gas will 
affect the SZ signal.
One can express
the strength of the integrated SZ flux as
\begin{equation} \label{eqn:lsz}
L_{SZ} = \int dA \int \rho {\rm k}T dl
\end{equation}
where the integration limits are along the line of sight
through the entire cluster, and over area 
out to projected radius $R_{500}$.
The right panel of Fig$.$\,\ref{fig:sz} displays
how the SZ signal varies with cluster mass in our model,
for the zero and maximum feedback cases.
At higher masses the relation 
has little dependence on feedback.
Feedback reduces the SZ signal somewhat (as it results
in gas being pushed out of the higher pressure cluster cores)
and makes the relation steeper.
Fitting to halos with $M_{500}\ge 10^{14}h^{-1}M_\odot$, we
find $L_{SZ}\propto M_{500}^{1.62}$ with zero feedback.
This is close to the self-similar slope of 5/3 predicted for 
spherical profiles \citep[e.g.][]{ReidSpergel06}; apparently
triaxiality and substructure have little effect
on this relation.
The slope steepens only slightly to
1.69 for $\epsilon_f=5\times 10^{-5}$;
this is in reasonable agreement with the analytic results
of \citet{ReidSpergel06} and 
with hydrodynamic 
simulations \citep{WhiteHS02,daSilvaKLT04,MotlHBN05,Nagai06}.
The exact slope found for this relation depends on the
lower mass limit used;  if we included lower mass halos
(or weighted the high mass halos less) the resulting slope
would be steeper.
The left panel of Fig$.$\,\ref{fig:sz} shows
how $L_{SZ}$ varies with temperature.  The
relation is quite tight, and again steepens with
increasing feedback; fitting to halos with ${\rm k}T_{sp}\ge 3$keV,
$L_{SZ}\propto T_{sp}^{2.43}$ with no feedback,
the exponent increasing to 
2.77 for $\epsilon_f=5\times 10^{-5}$.
The zero feedback slope agrees well with the adiabatic
simulation of \citet{Nagai06}, but the feedback
model is steeper, because we are putting
in more energy.  Increasing $\epsilon_f$ from zero
to $\epsilon_f=5\times 10^{-5}$ lowers the SZ signal at
k$T_{ew}=5$keV by 35\%, which is similar to but
slightly less than the effect seen by \citet{Nagai06}
between his adiabatic and star formation runs.   A
more detailed comparison is difficult because we
are using the projected SZ signal; also there are
differences in the cosmological parameters.

This work makes it clear that allowance for feedback will be
necessary if one is to utilize the upcoming SZ surveys
for precision measurements of cosmological parameters.
Fortunately, X-ray observations allow us to calibrate
the feedback parameter; adequate fits to the data can
be obtained if the ratio of energy input to stellar
mass is $\epsilon_f=3-5\times 10^{-5}$. 
Uncertainties in this
parameter will propagate into uncertainties in the mass-flux decrement
relation for SZ surveys.  However, it can be seen in Fig$.$\,\ref{fig:sz}
that even the extreme case of reducing $\epsilon_f$ to zero
hardly changes this relation for clusters
with masses above $2\times 10^{14}h^{-1}M_\odot$.  Moreover, it is not the
scatter in the mass-observable relation which makes it difficult for
future SZ surveys to constrain cosmological parameters, 
but rather it is the
uncertainty in the scatter which is the 
main problem \citep{Lima:05}.  
We will use the methods of this paper to explore these issues
in future work.


\acknowledgments

The authors would like to thank  Niayesh Afshordi for
useful email exchanges,
and Ian McCarthy for making available the observational data.
We also thank the anonymous referee for a careful and helpful 
reading of the manuscript.
This work was partially supported by the National Center for
Supercomputing Applications under grant MCA04N002; 
in addition, computational facilities at Princeton supported 
by NSF grant AST-0216105 were used, as well as
high performance computational facilities supported by 
Princeton University under the auspices of the
Princeton Institute for Computational Science and Engineering 
(PICSciE) and the Office of Information Technology (OIT).

\clearpage


\epsscale{0.90}
\epsscale{1.0}

%
\begin{deluxetable}{lcclcc}
\tabletypesize{\scriptsize}
\tablecaption{Test of Resolution for low-mass Halos\label{tab:restest}}
\tablewidth{0pt}
\tablehead{
\colhead{ }                                      &
\multicolumn{2}{c}{$1< M_{500} \le 4$} &
\colhead{ }                                      &
\multicolumn{2}{c}{$4< M_{500} \le 6$} \\
\colhead{ }                                      &
\colhead{Standard}                                    &
\colhead{High res.}                                   &
\colhead{ }                                      &
\colhead{Standard}                                    &
\colhead{High res.}                                   }
\startdata
$M_{500}$  & $3.18 \pm 0.55$ & $3.29 \pm 0.51$ & & 
  4$.85 \pm 0.57$ & $4.81 \pm 0.56$ \\
${\rm k}T_{ew}$ & $1.68 \pm 0.19$  & 1$.75 \pm 0.21$ & & 
  $1.99 \pm 0.24$ & $1.97 \pm 0.21$ \\
$L_x$  & $1.14 \pm 0.76$ & $1.21 \pm 1.02$ & & 
  $3.63 \pm 1.99$ & $3.79 \pm 2.77$  \\
$L_{SZ}$  & $3.31 \pm 1.95$ & $3.16 \pm 1.95$ & & 
  $7.13 \pm 3.91$ & $6.24 \pm 2.94$
\enddata
\tablecomments{Mean and one standard deviation.
Units of $M_{500}$, ${\rm k}T_{ew}$, $L_x$, $L_{SZ}$: 
$10^{13}h^{-1}M_\odot$, keV, $10^{42}h^{-2}{\rm erg}\,s^{-1}$, 
$10^{12} h^{-1}$keV$M_\odot$ }
\end{deluxetable}
\begin{deluxetable}{ccclcc}
\tabletypesize{\scriptsize}
\tablecaption{Power Law Fit Parameters \label{tab:olsfits}}
\tablewidth{0pt}
\tablehead{
\colhead{ }                                 &
\multicolumn{2}{c}{$\epsilon_f=0$}              &
\colhead{ }                                 &
\multicolumn{2}{c}{$\epsilon_f=5\times10^{-5}$} \\
\colhead{Relation}                          &
\colhead{$A$}                                 &
\colhead{$\alpha$}                           &
\colhead{ }                                 &
\colhead{$A$}                                 &
\colhead{$\alpha$}                           }
\startdata
${\rm k}T_{sp} - M_{500}$ & 0.46 $\pm$ 9\% & 0.67 $\pm$ 0.01 &  & 0.55 $\pm$ 10\% & 0.60 $\pm$ 0.01 \\
$M_g - M_{500}$           & 1.44 $\pm$ 10\% & 0.96 $\pm$ 0.01 &  & 0.80 $\pm$ 12\% & 1.17 $\pm$ 0.02 \\
$L_x - M_{500}$           & 1.15 $\pm$ 26\% & 1.13 $\pm$ 0.02 &  & 0.31 $\pm$ 25\% & 1.59 $\pm$ 0.03 \\
$L_{SZ} - M_{500}$        & 0.43 $\pm$ 25\% & 1.62 $\pm$ 0.02 &  & 0.35 $\pm$ 31\% & 1.69 $\pm$ 0.03 \\
$M_{500} - {\rm k}T_{sp}$ & 3.12 $\pm$ 14\% & 1.49 $\pm$ 0.02 &  & 2.56 $\pm$ 19\% & 1.62 $\pm$ 0.03 \\
$M_g - {\rm k}T_{sp}$     & 4.37 $\pm$ 17\% & 1.40 $\pm$ 0.03 &  & 2.42 $\pm$ 24\% & 1.84 $\pm$ 0.06 \\
$L_x - {\rm k}T_{sp}$     & 4.09 $\pm$ 29\% & 1.66 $\pm$ 0.03 &  & 1.36 $\pm$ 34\% & 2.51 $\pm$ 0.08 \\
$L_{SZ} - {\rm k}T_{sp}$  & 2.86 $\pm$ 15\% & 2.43 $\pm$ 0.03 &  & 1.85 $\pm$ 16\% & 2.77 $\pm$ 0.04 \\
\enddata
\tablecomments{ For each relation $Y-X$, the best fit parameters
of the form $Y/Y_0=A(X/X_0)^\alpha$ are given.  For
$Y=( M_{500}, {\rm k}T_{sp}, M_g, L_x, L_{SZ}), 
Y_0=( 10^{14}h^{-1}M_\odot, 5{\rm keV}, 10^{13}h^{-1}M_\odot, 
10^{44}h^{-2}{\rm erg}\,s^{-1}, 10^{14}h^{-1}{\rm keV}M_\odot)$.
When $X= M_{500}$, the fit is to all halos with 
$X\ge X_0=10^{14}h^{-1}M_\odot$; when $X={\rm k}T_{sp}$, 
$X_0=5$keV and the fit
is to all halos with ${\rm k}T_{sp}\ge 3$keV.
The scatter given for $A$  is the $rms$ fractional
difference of the halos from the best-fit relation.
}
\end{deluxetable}

\end{document}